\documentclass[]{article}
\usepackage{amsmath,amsfonts}
\usepackage{algorithmic}
\usepackage{algorithm}
\usepackage{array}
\usepackage[caption=false,font=normalsize,labelfont=sf,textfont=sf]{subfig}
\usepackage[
    paperwidth=8.5in,
    paperheight=11in,
    left=1in,
    right=1in,
    top=0.5in,
    bottom=0.75in,
    headheight=15pt,   % plays nicer with fancyhdr
    headsep=0.05in
]{geometry}

\usepackage{textcomp}
\usepackage{stfloats}
\usepackage{url}
\usepackage{verbatim}
\usepackage{graphicx}
\usepackage{cite}
\usepackage{caption}
\usepackage{xcolor}

\usepackage{fancyhdr}
\fancypagestyle{firstpage}
{
  \fancyhf{}
  \fancyfoot[C]{\footnotesize \copyright 2025 IEEE. All rights reserved, including rights for text and data mining, and training of artificial intelligence and similar technologies. Personal use is permitted, but republication/redistribution requires IEEE permission. See https://www.ieee.org/publications/rights/index.html for more information. \textit{DOI: 10.1109/TDMR.2025.3576042}}
}
\begin{document}

\title{The First Switch Effect in Ferroelectric Field-Effect Transistors\\ 
\author{}
\vspace{2mm}

\small{{Priyankka Ravikumar$^{1,\wedge}$, Prasanna Venkatesan$^1$, Chinsung Park$^1$, Nashrah Afroze$^1$\\ Mengkun Tian$^2$, Winston Chern$^1$, Suman Datta$^{1,3}$, Shimeng Yu$^1$, Souvik Mahapatra$^{4}$, Asif Khan$^{1,3,\$}$ \\
$^1$School of Electrical and Computer Engineering, Georgia Institute of Technology, GA, USA; $^2$Institute of Materials, GA, USA; $^3$School of Material Science and Engineering, Georgia Institute of Technology, GA, USA; $^4$Department of Electrical Engineering, Indian Institute of Technology, Bombay, Mumbai, India. \\$\{${$^{\wedge}$pravikumar30, $^{\$}$akhan40}$\}$@gatech.edu}}}
\date{}
\vspace{0.75in}

% The paper headers
\markboth{}%
{Shell \MakeLowercase{\textit{et al.}}: A Sample Article Using IEEEtran.cls for IEEE Journals}

\maketitle
\thispagestyle{firstpage}

\begin{abstract}
In this work, a ferroelectric field-effect transistor (FEFET) is systematically characterized and compared with an equivalent standard MOSFET with an equivalent oxide thickness. We show that these two devices, with a silicon channel, exhibit similar pristine state transfer characteristics but starkly different endurance characteristics. In contrast to the MOSFET, the FEFET shows a significant increase in sub-threshold swing in the first write pulse. Based on this, we reveal that this first write pulse (cycle 1) generates more than half of the total traps generated during the fatigue cycling in FEFETs. We call this the ``First Switch Effect". Further, by polarizing a pristine FEFET step by step, we demonstrate a direct correlation between the switched polarization and interface trap density during the first switch. Through charge pumping measurements, we also observe that continued cycling generates traps more towards the bulk of the stack, away from the Si/SiO$_2$ interface in FEFETs. We establish that: (1) the first switch effect leads to approximately 50$\%$ of the total trap density (N$_{it}$) near the Si/SiO$_2$ interface until memory window closure; and (2) further bipolar cycling leads to trap generation both at and away from Si/SiO$_2$ interface in FEFETs.
\\
\newline
HfO$_2$-based Ferroelectrics, FEFET, Endurance, Charge pumping.
\end{abstract}
\section{\textbf{Introduction}}

Ferroelectric field-effect transistors (FEFETs) are gaining attention as a next-generation non-volatile memory solution due to their high switching speed, high density, and extremely low power consumption\cite{intro_1,intro_2}. These characteristics make them particularly attractive for applications in artificial intelligence, edge computing, and large-scale cloud storage. However, a major challenge limiting their widespread deployment is their restricted write endurance, primarily indicated by memory window (MW) closure under repeated bipolar stress. While MW degradation has been correlated with polarization switching\cite{1,2,3,4,5,intro_3}, the fundamental mechanisms driving charge trapping and its contribution to device failure remain unclear. Gaining deeper insights into interface states and trap formation processes in FEFETs is essential to enhancing their endurance and ensuring their long-term reliability.\\

This work introduces a lesser-known degradation mechanism in FEFETs, termed the \textbf{first switch effect}. We demonstrate that the very first write pulse applied to a pristine (as-fabricated) device induces considerable interfacial damage, accounting for almost half of the total device degradation before MW closure, which typically occurs between \(10^4\)–\(10^5\) cycles. Although repeated cycling further contributes to degradation, our findings suggest that FEFET endurance deterioration occurs in two distinct phases:

\noindent 1) First switch-induced degradation, where the initial pulse significantly impacts the interface.  
2) Degradation due to bipolar cycling, where continued stress gradually leads to MW closure (Fig. 1). \\

Our results indicate that significant degradation occurs as soon as the first voltage pulse is applied, as reflected by the noticeable increase in subthreshold swing immediately after this event (Fig. 5(d)). The first switch effect arises due to the initial random alignment of ferroelectric dipoles in a pristine device. When an external electric field is applied, the dipoles reorient along the field direction, exposing the interfacial layer (IL) to a strong electric field due to unscreened polarization charges. This exposure results in extensive trap formation within the IL (Fig. 1).  To substantiate this finding, we employ frequency-dependent charge pumping measurements, which confirm that trap generation in FEFETs occurs both near and deeper within the Si/SiO\(_2\) interface, unlike MOSFETs, where trap generation is mainly localized near the interface.

 \begin{figure}[!t]
    \centering
    \includegraphics[width=0.7\textwidth]{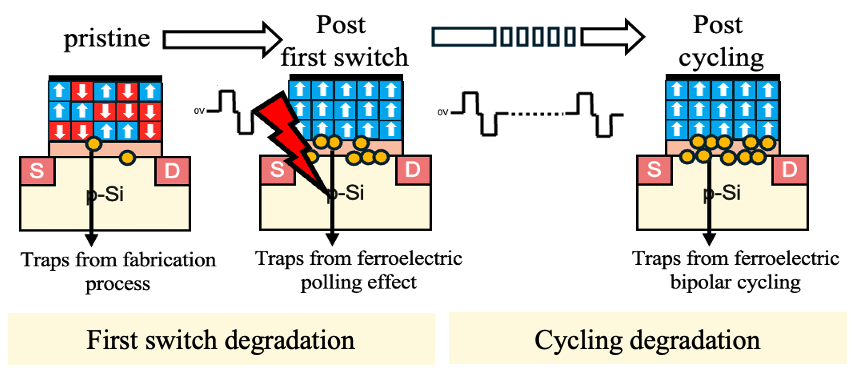}
    \caption{Schematic showing the first switch and cycling degradation stages in FEFETs.}
    \label{fig:enter-label1}
\end{figure}

 \begin{figure}[!b]
   \centering
    \includegraphics[width=0.7\textwidth]{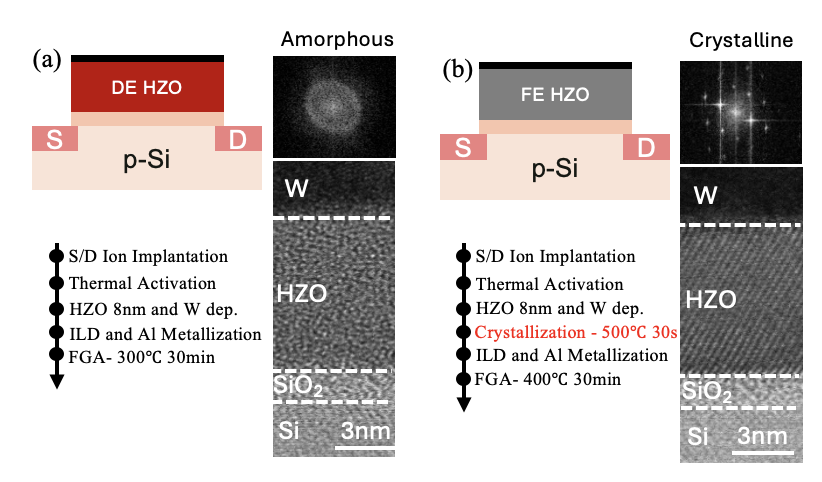}
    \caption{Process flow and STEM image of crystalline FE-HZO FET and amorphous dielectric HZO FET. Diffraction patterns confirm the crystallinity and amorphous nature of FEFET and MOSFET respectively. }
    \label{fig:enter-label2}
\end{figure}

\section{\textbf{Fabrication process and basic characterization}}

{n-channel FEFETs were fabricated with a gate stack of 8 nm of $Hf_{0.5} Zr_{0.5} O_{2}$ as the ferroelectric layer and 1 nm of silicon dioxide ($SiO_2$) as the interfacial layer on a p type Si wafer. n+ source and drain were formed using ion implantation followed by dopant activation. The HZO film is deposited at 250$^\circ$C using thermal ALD followed by crystallization via a rapid thermal anneal with a tungsten (W) capping layer at 500$^\circ$C for 30 seconds This is followed by a forming gas anneal (FGA) at 400$^\circ$C for 30 mins. The first anneal, Rapid thermal anneal, is performed to crystallize the HZO layer. The second anneal, forming gas anneal is performed to improve the quality of the $Si-SiO_{2}$. For further details in the fabrication procedure, refer to previous publications \cite{CP_T_P}. An equivalent MOSFET was also fabricated using the same procedure as the FEFET but without the rapid thermal anneal step resulting in an amorphous dielectric HZO layer instead of a ferroelectric HZO layer. The device schematic is shown in Fig. 2. The gate length, L$_G$, and width, W, of the measured device in this paper is 3 $\mu$m and 10 $\mu$m, respectively. Transmission electron microscopy (TEM) imaging and the corresponding diffraction pattern of the FEFET and the MOSFET in Fig. 2 show the crystallinity of the ferroelectric layer in the FEFET and the amorphous nature of HZO in the MOSFET.

\vspace{0.1in}

The ferroelectric nature of the 8 nm FEFET was confirmed from the hysteretic P-V loop  while the MOSFET exhibited a linear dielectric response (Fig. 3(c,d)). These two devices have the same effective oxide thickness (EOT), as seen from the C-V curves measured at a frequency of 1MHz in Fig. 3(a,b), allowing for a direct comparison of the two devices. The transfer characteristics and MW-$V_{write}$ characteristics of the two devices show no MW in the case of the MOSFET and a max MW of 1.5 V @ $V_{write}$ = 3.5 V in the case of FEFET. The MW is measured by (Fig. 5 (a), bipolar) first applying a pre-polarization pulse of pulse width 10 $\mu s$ followed by a erase pulse of pulse width 1 $\mu s$. This pulse sets the device to the high V$_T$ state and is read using a voltage sweep from -0.5 V to 2 V. Subsequently the program pulse of 1 $\mu s$ is applied and the low V$_T$ state is read again. The MW is defined by a constant current line of 0.1 $\mu$A/$\mu m$ in the I$_d$-V$_g$ curves It can also be noted from the transfer characteristics that the MOSFET has a lower subthreshold swing compared to the FEFET.

\begin{figure}[!t]
    \centering
    \includegraphics[width=0.5\textwidth]{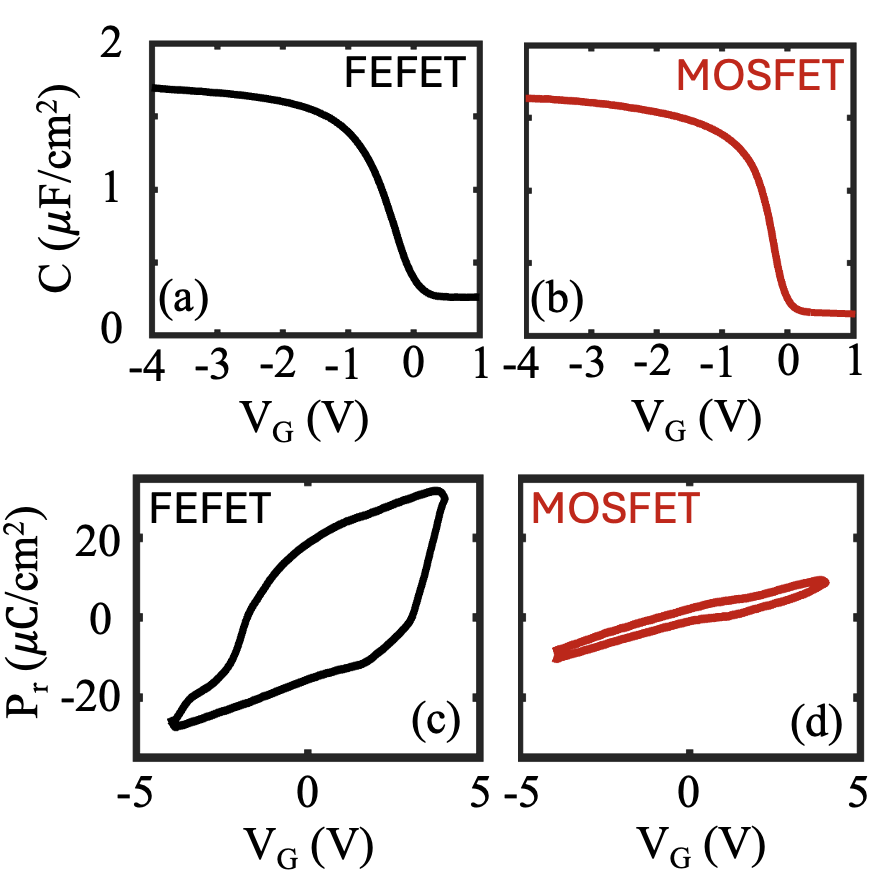}
    \caption{a,b) C-V curves showing the same  effective oxide thickness (EOT) for both devices c,d) PV loops measured for the MOSFET and FEFET. The FEFET shows a clear hysteresis loop while the MOSFET shows a dielectric response. }
    \label{fig:enter-label3}
\end{figure}

\begin{figure}[!t]
    \centering
    \includegraphics[scale=0.55]{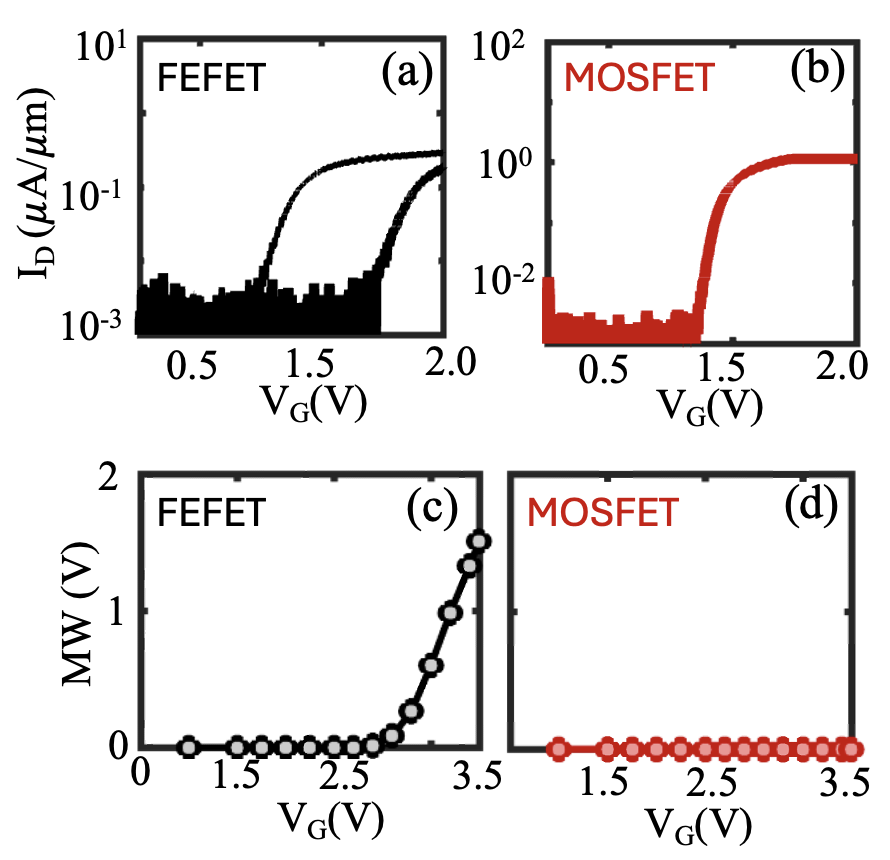}
    \caption{a,b) I$_d$-V$_g$ curves for the MOSFET and FEFET. b,c) Memory window vs voltage curves for the MOSFET and FEFET. The FEFET shows a maximum memory window of 1.5 V  at a program voltage of 3.5 V while the MOSFET does not show any memory window.}
    \label{fig:enter-label4}
\end{figure}

\section{\textbf{Experiments and Discussion}}

\begin{figure*}[!h]
    \centering
    \includegraphics[scale=0.8]{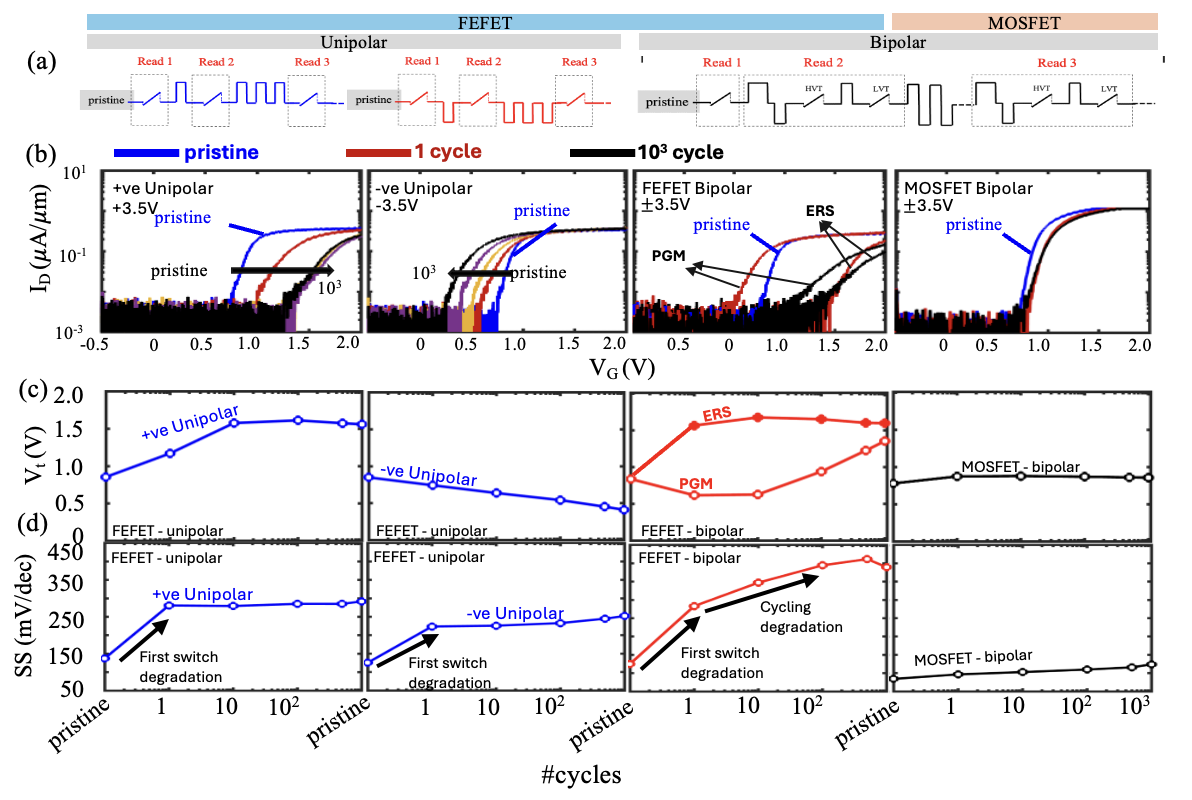}
    \caption{Measurement scheme for unipolar and bipolar measurements in these devices. b) $I_d-V_g$ for the FEFET under bipolar stress, unipolar stress and MOSFET under bipolar stress respectively. c) $V_T$ shift with cycles for pristine and post first switch cases. d) Change in subthreshold swing with cycles.}
    \label{fig:enter-label5}
\end{figure*}

We investigate the behavior of ferroelectric field-effect transistors (FEFETs) under three distinct electrical stress conditions to gain deeper insights into the interplay between polarization switching and trap dynamics. The three stress conditions include: (1) unipolar cycling with a positive voltage of +3.5 V, (2) unipolar cycling with a negative voltage of -3.5 V, and (3) bipolar cycling with alternating voltage pulses of ±3.5 V. These stress conditions are applied starting from the pristine state and continuing up to 1000 cycles, as illustrated in Fig. 5(a). The chosen stress voltage is derived from the measured maximum memory window ($MW$) in the $MW-V_{write}$ characteristics of the FEFET (Fig 4(c)). To monitor the impact of these stress conditions on the device performance, we measure the pulsed $I_d-V_g$ characteristics at regular intervals throughout the cycling process. This enables us to track the evolution of subthreshold swing and the threshold voltage ($V_T$) under different stress conditions, as shown in Fig. 5(b). The use of both unipolar and bipolar stress conditions allows us to decouple the individual contributions of polarization switching and voltage stress on the device degradation. By examining the effects of unipolar stress separately for positive and negative voltages, we can distinguish its influence on the program (PGM) state, which corresponds to the low-$V_T$ condition, and the erase (ERS) state, which corresponds to the high-$V_T$ condition of the FEFET. Under positive unipolar stress, we observe a gradual positive shift in $V_T$, whereas under negative unipolar stress, $V_T$ shifts in the negative direction. In contrast, during bipolar cycling, the $V_T$ values of both the PGM and ERS states progressively converge. This convergence leads to a gradual reduction of the MW, eventually resulting in MW closure after approximately 5000 cycles. \\

In the pristine state, before any write pulse is applied, the subthreshold slope of the FEFET is measured to be 120 mV/dec. However, as shown in Fig. 5(d), the application of the very first write pulse—regardless of whether it is part of a unipolar or bipolar stress pulse train—induces a significant and immediate increase in SS, jumping from 120 mV/dec to 280 mV/dec. We call this significant degradation the \textit{``first switch effect"}. \\

Further unipolar cycling, regardless of whether it involves positive or negative voltage stress, is observed to result in minimal to no additional change in SS beyond the initial increase caused by the first switch. This indicates that, under unipolar stress conditions, the primary degradation in SS occurs almost entirely during the first write event, with subsequent cycling having little to no further impact on the device’s subthreshold characteristics. In contrast, under bipolar cycling, the SS continues to degrade progressively beyond the first switch event, with a steady increase until memory window (MW) closure is reached. Specifically, the SS increases from its post-first-switch value of 280 mV/dec to a final value of 450 mV/dec by the time MW closure occurs. Notably, nearly half of the total SS degradation observed under bipolar cycling—amounting to a significant portion of the overall increase—takes place solely during the very first write pulse. The continued degradation of the SS after the first switch can be attributed to polarization switching during the bipolar stress pulses. On the contrary, when a standard MOSFET is subjected to the same bipolar stress conditions, only minimal changes in both threshold voltage and SS  are observed. While there is a minor shift in the threshold voltage between the pristine state and the first switched state in the MOSFET, this can be attributed to electron trapping due to the first cycling pulse. This is in conjunction with a slight degradation in SS that is observed as well. In this case, the SS increases from 80 to 100 mV/dec and $V_{T}$ shifts by 0.1 V after the first switch. However, this degradation is significantly lower than the degradation observed in FEFETs for the same applied voltage pulses. This stark difference in behavior between the standard MOSFET and the FEFET strongly suggests that the first switch effect, as well as the progressive shifts in $V_T$, are inherently linked to ferroelectric switching. 

\vspace{0.1in}

\begin{figure*}[!t]
    \centering
    \includegraphics[scale=0.6]{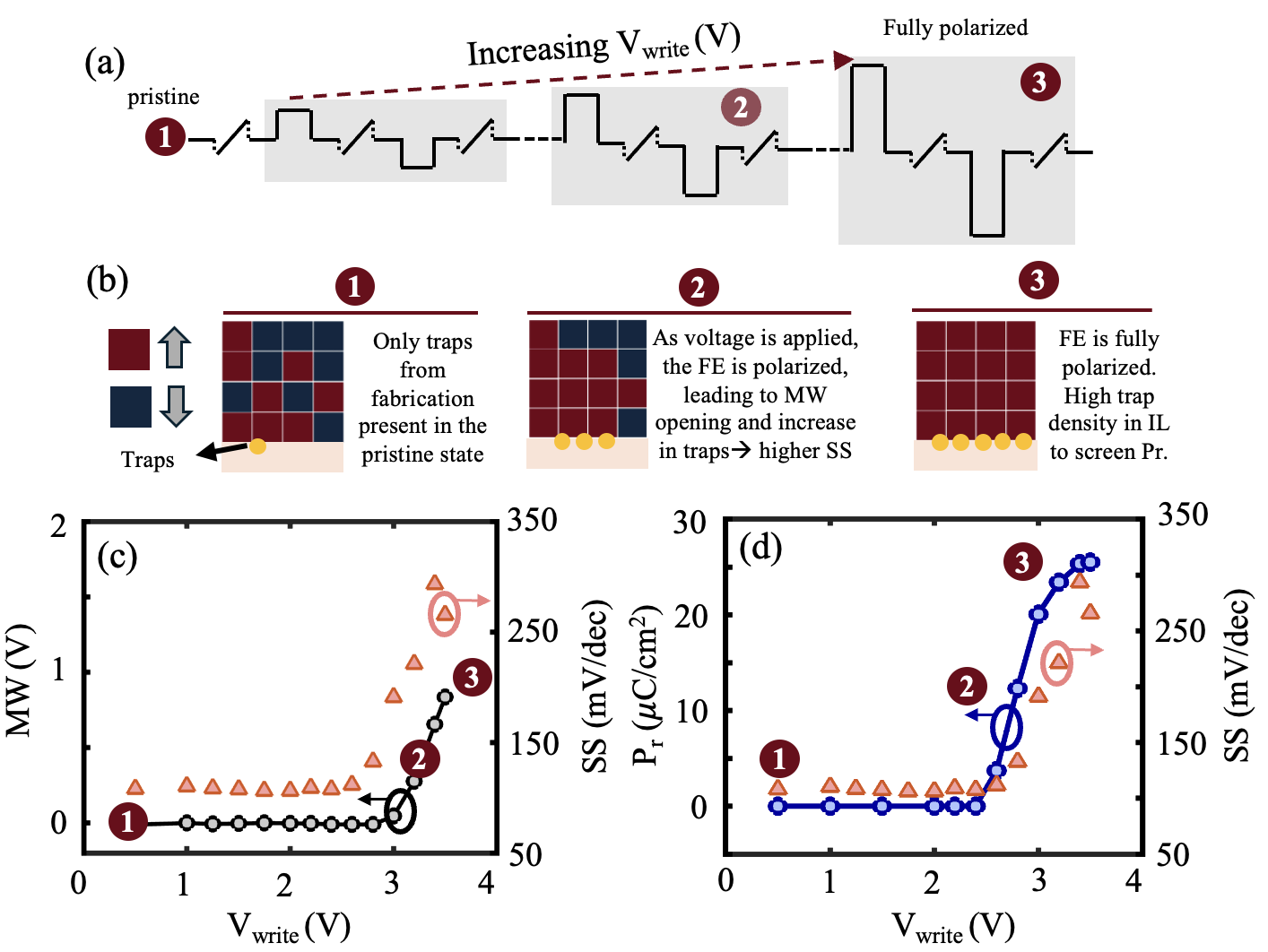}
    \caption{ a) Measurement scheme to polarize a pristine FEFET step by step and track the memory window and the SS. b)Schematic of the different stages of voltage application with initial stage showing low Pr and low trap density and final stage showing higher Pr and high trap density. c)MW vs Voltage curve for the performed experiment and the corresponding SS at each applied voltage. d) Pr vs Voltage measured using PUND (Positive-Up-Negative-Down) measurements and the corresponding SS.}
    \label{fig:enter-label6}
\end{figure*}
\begin{figure}[!t]
    \centering
    \includegraphics[scale=0.435]{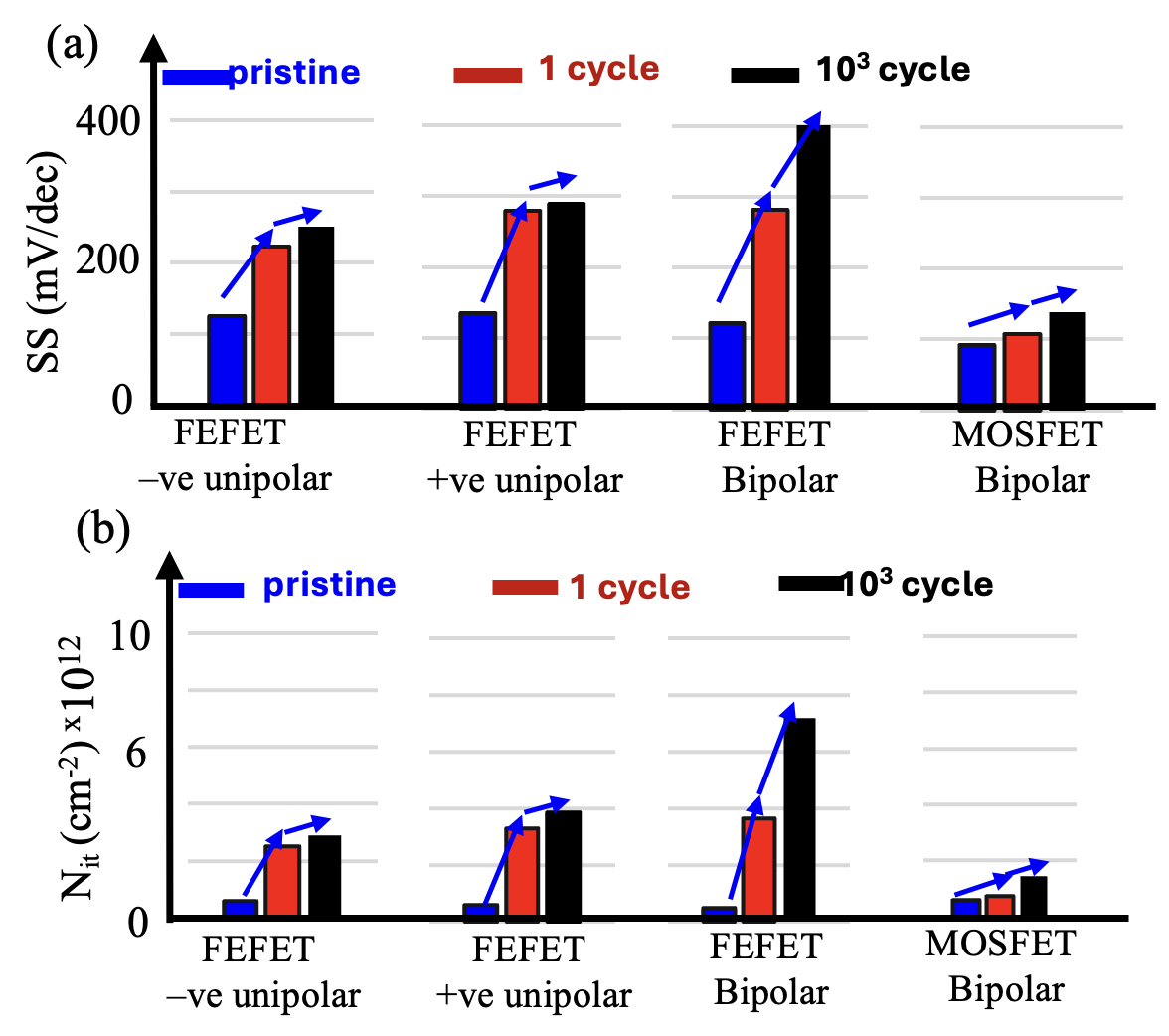}
    \caption{(\small{a) Bar charts show the amount of subthreshold swing degradation in FEFETs and MOSFET under different stress conditions. b) Extracted $N_{it}$ from subthreshold swing.} }
    \label{fig:enter-label7}
\end{figure}

Figure 7 provides a comprehensive summary of the SS and the corresponding defect densities extracted from SS measurements. A key observation from the data is that the first switch effect is responsible for nearly 90$\%$  of the total defect generation in the case of unipolar cycling. This indicates that the majority of defects are introduced during the very first write pulse, with subsequent unipolar cycling contributing minimally to further defect formation. In contrast, under bipolar cycling, the first switch effect accounts for approximately 50$\%$ of the total defect generation until memory window (MW) closure. This suggests that while the initial write pulse introduces a significant number of defects, the remaining defect generation is progressively induced by repeated polarization switching during continued bipolar stress.\\

To further investigate the first switch effect in greater detail, we conduct an experiment using a pristine FEFET, applying voltage pulses of gradually increasing magnitude. After each applied pulse, an I$_d$-V$_g$ sweep is performed to measure the resulting memory window (MW), allowing us to track how the MW evolves as the pulse amplitude increases. The pulse scheme used for this experiment is illustrated in Fig. 6(a). By progressively increasing the pulse amplitude, we systematically observe the relationship between memory window and device degradation. As shown in Fig. 6(c), the measured MW increases as the pulse amplitude increases. Notably, the plot also reveals a concurrent increase in SS, indicating that as the MW opens up, additional trap states are introduced into the IL. This suggests a direct correlation between polarization switching and trap generation. Fig. 6(d) presents the remnant polarization extracted from PUND (Positive-Up-Negative-Down) measurements, providing further insight into the correlation between polarization switched and SS. The results show that as more polarization is switched, an increasing number of traps are generated within the interfacial layer. Fig 6(b) demonstrates the first switch effect, showing that as more of the ferroelectric layer is polarized, the SS and hence the trap density in the IL also increases. This correlation between the polarization switched and the increase in SS is likely due to a strong electric field seen by the IL layer due to unscreened polarization charges. As polarization increases, the electric field across the IL increases, leading to a higher number of traps generated. 

\vspace{0.1in}

To gain a deeper understanding of the traps generated by the first switch effect and subsequent bipolar cycling, charge pumping measurements were conducted on both the FEFET and the standard MOSFET. During the charge pumping process, voltage pulses with fixed amplitude but varying base voltages ($V_{base}$) are applied to the gate. The source and body contacts are shorted and grounded and the base is used to collect the recombination current. This technique enables the scanning of the band gap by cycling the device through the accumulation and inversion states.
The device is initiated by flooding the traps with holes. When the higher voltage inverts the channel, electrons come in and recombine with the holes, resulting in a recombination current that can be measured at the body (Fig 8(a)). \\

 The charge pumping measurements were performed at 1 MHz, 100 kHz and 20 kHz with the rise and fall times of the pulse kept constant. High-frequency measurements detect the recombination current of fast traps near the channel, whereas lower frequency measurements capture the contribution of traps deeper into the gate stack. The trap density extracted for each of the frequencies from the recombination current is given by 
 
\begin{equation}
I_{cp} = q A f N_{it}
\end{equation}

where $q$ is electronic charge and $A$ is the area of the device \cite{charge_pumping}. The trap density measured using charge pumping increases with cycling in both the MOSFET and the FEFET. However, in the FEFET, as the frequency is lowered, the trap density drastically increases, indicating that the traps being generated are present near and away from the channel of the FEFET. In the MOSFET, the lowering of the frequency does not have a significant impact on the magnitude of trap density measured, indicating that trap generation occurs primarily near the channel of the MOSFET (Fig.8 (b)).

 \begin{figure}[!t]
    \centering
    \includegraphics[scale=0.43]{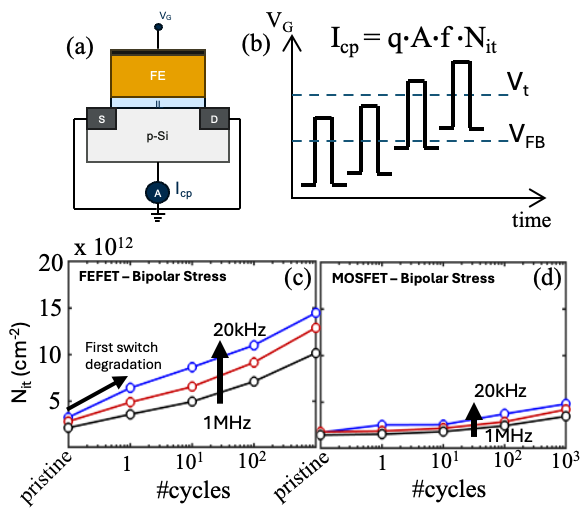}
    \caption{ a, b) Schematic of charge pumping measurement setup: the device goes from accumulation to inversion repeatedly while the body current is measured to extract the recombination current. c, d) High and low frequency charge pumping on the FEFET and MOSFET. FEFET shows degradation near channel interface and deeper in the stack while the MOSFET degradation is predominantly at the Si-SiO$_2$ interface.}
    \label{fig:enter-label8}
\end{figure}

\section{\textbf{Conclusion}}
While the degradation of ferroelectric field-effect transistors (FEFETs) due to bipolar cycling is commonly explored, the impact of the initial polarization switch within the ferroelectric layer remains largely unexplored. When a voltage is first applied to the gate of a pristine FEFET, polarization domains align in one direction, subjecting the interfacial layer (IL) to a strong electric field due to unscreened polarization charges. This initial switching event leads to a more significant degradation in the subthreshold slope (SS) compared to the subsequent 1000 bipolar stress cycles, accounting for nearly 50$\%$ of the total SS degradation before memory window closure. Slowly increasing the polarization of the ferroelectric layer using voltage pulses reveal a direct correlation between the amount of polarization switched and the defect density in the IL during the first switch. Further, charge pumping measurements show that cycling the FEFET causes damage not only at the SiO$_2$-Si interface but also deeper within the gate stack. In contrast, MOSFET degradation is primarily confined to the SiO$_2$-Si interface. Therefore, mitigating the first switch effect in FEFETs could play a crucial role in enhancing the overall endurance of ferroelectric field effect transistors.

\section*{\textbf{Acknowledgement}}
This work was supported by SUPREME, one of the seven SRC-DARPA JUMP2.0 centers. Fab was done at the IEN, supported by the NSF-NNCI program (ECCS- 1542174).

\vfill

\end{document}